\documentstyle[12pt]{article}
\begin{document}
\centerline{\Large\bf Noether Charge and Black Hole Entropy}
\vskip 0.1in
\centerline{\Large\bf in Modified Theories of Gravity}
\vskip .7in
\centerline{Dan N. Vollick}
\vskip .1in
\centerline{Irving K. Barber School of Arts and Sciences}
\centerline{University of British Columbia Okanagan}
\centerline{3333 University Way}
\centerline{Kelowna, B.C.}
\centerline{Canada}
\centerline{V1V 1V7}
\vskip .1in
\centerline{and}
\vskip .1in
\centerline{Pacific Institute for Theoretical Physics}
\centerline{Department of Physics and Astronomy}
\centerline{University of British Columbia}
\centerline{6224 Agricultural Road}
\centerline{Vancouver, B.C.}
\centerline{Canada}
\centerline{V6T 1Z1}
\vskip .9in
\centerline{\bf\large Abstract}
\vskip 0.5in
\noindent

The entropy of black holes in modified theories of gravity is examined in the Palatini formalism using the Noether Charge approach.
It is shown that, if the gravitational coupling constant is properly
identified, the entropy of a black hole is one-quarter of the horizon area in f(R) theories coupled to conformally invariant matter. If
matter is present that is not conformally invariant the entropy is still proportional to the area
of the black hole, but the coefficient is generally not one-quarter.
The entropy of black holes in generalized dilaton theories and in theories with Lagrangians that
depend on an arbitrary function of the Ricci tensor are also examined.
\newpage
\section{Introduction}
Over the last few years there has been a significant amount of interest in modifications
of gravity as an explanation for the observed accelerated expansion of the
Universe \cite{1} (see also \cite{others 1}). There are two approaches to derive
the field equations from the action in these theories, the metric approach and the Palatini
approach. In the metric approach the connection is taken to be the Christoffel
symbol and the action is varied with respect to the metric. In the Palatini
approach the metric and connection are varied independently. For the action
of general relativity these approaches produce the same field equations. However,
for other actions they produce different theories.

In this paper I examine the entropy of black holes in modified theories of
gravity in the Palatini formalism using the Noether charge approach as
developed by Wald \cite{Wa1} (see also
\cite{Wa2,Vi1,Ja1,Wa3}).
In this approach to black hole entropy the first law of black hole mechanics
is written as
\begin{equation}
\delta\int_{\Sigma}{\bf Q}=
\delta E-\Omega_{H}^{(A)}\delta J_{(A)}\; ,
\end{equation}
where ${\bf Q}$ is the Noether potential associated with diffeomorphisms
on the manifold, $\Sigma$ is the bifurcation surface of the black hole,
$E$ is the canonical energy, $J_{(A)}$ is the canonical angular
momentum and $\Omega_{(A)}$ is the angular velocity of the horizon.
If ${\bf Q}$
can be written in terms of local geometric quantities and the matter
fields present in the spacetime the entropy will be given by
\begin{equation}
S_{BH}=\frac{2\pi}{\kappa}\int_{\Sigma}{\bf Q}\; ,
\end{equation}
where $\kappa$ is the surface gravity of the black hole. In fact, it has
been shown \cite{Ja1} that the above integral can be taken over any
cross-section of the Killing horizon.

In section 2 the Noether charge approach is applied to gravitational theories
in the Palatini formalism. In section 3 the entropy of black holes in f(R)
theories of gravity is investigated. In section 4 theories with Lagrangians that
depend on arbitrary functions of the Ricci tensor and generalized dilaton theories are examined.

\section{Noether Charge and Black Hole Entropy}
In this section I will apply the Noether charge approach, developed
by Wald \cite{Wa1} in the metric formalism, to calculate black hole entropy in the Palatini formalism
(see (\cite{Br1}) for work on black hole entropy in the metric and other approaches).
The Lagrangian D-form in D dimensions will be taken to be
\begin{equation}
{\bf L}=\frac{1}{16\pi G}f(g^{\mu\nu},R^{\alpha}_{\;\;\mu\nu\lambda},\psi,.
. .){\bf \epsilon}\; ,
\end{equation}
where
${\bf \epsilon}$ is the volume form on the manifold,
\begin{equation}
R^{\alpha}_{\;\;\mu\nu\lambda}=\partial_{\nu}\Gamma^{\alpha}_{\mu\lambda}-
\partial_{\lambda}\Gamma^{\alpha}_{\mu\nu}+\Gamma^{\alpha}_{\nu\sigma}
\Gamma^{\sigma}_{\mu\lambda}-\Gamma^{\alpha}_{\lambda\sigma}\Gamma^{\sigma}
_{\mu\nu}\; ,
\end{equation}
$\psi$ denotes the matter fields and ... denotes derivatives of $g^{\mu\nu}$ and
$R^{\alpha}_{\;\;\mu\nu\lambda}$. Note that $g^{\mu\nu}$ and $\Gamma^{\alpha}_
{\mu\nu}$ are to be taken as independent field variables and that
$\Gamma^{\alpha}_{\mu\nu}$ is taken to be symmetric in $\mu$ and $\nu$. Let
$\phi=(g^{\mu\nu},\Gamma^{\alpha}_{\mu\nu},\psi)$ denote the field
variables. The variation in the Lagrangian associated with the variation in the field variables,
$\delta\phi$, is given by
\begin{equation}
\delta {\bf L}={\bf E}\cdot\delta\phi+{\bf d\Theta}(\phi,\delta
\phi)\; ,
\end{equation}
where ${\bf E}\cdot\delta\phi$ implies a summation over field and spacetime indices. The Noether current, ${\bf J}$, associated with the diffeomorphisms generated by the vector field $\xi^{\mu}$ is defined to be
\begin{equation}
{\bf J}={\bf \Theta}(\phi,L_{\xi}\phi)-{\bf L}\xi^{\mu}\; ,
\label{current}
\end{equation}
where $L_{\xi}\phi$ is the Lie derivative of $\phi$ with respect to $\xi^{\mu}$. At this point one might be concerned by the fact that $\Gamma^{\alpha}_{\mu\nu}$ is not a tensor
(although $\delta\Gamma^{\alpha}_{\mu\nu}$ is a tensor). However, defining
\begin{equation}
\left(L_{\xi}\Gamma\right)^{\alpha}_{\mu\nu}(x)=\lim_{\lambda\rightarrow 0}\left[
\frac{\bar{\Gamma}^{\alpha}_{\mu\nu}(x)-\Gamma^{\alpha}_{\mu\nu}(x)}{\lambda}
\right]
\end{equation}
with $\bar{x}^{\mu}=x^{\mu}-\lambda\xi^{\mu}$ gives the covariant expression
\begin{equation}
\left(L_{\xi}\Gamma\right)^{\alpha}_{\mu\nu}=\nabla_{\mu}\nabla_{\nu}\xi^
{\alpha}-R^{\alpha}_{\;\;\nu\mu\rho}\xi^{\rho}\; .
\label{Gamma}
\end{equation}
When the field equations are satisfied (i.e. when ${\bf E}=0$) it can be shown
that ${\bf dJ}=0$ for all $\xi^{\mu}$.
Thus, in this case, there exists a Noether potential, ${\bf Q}$, which satisfies
${\bf J}={\bf dQ}$ \cite{Wa4}.
The Noether potential ${\bf Q}$ is constructed out of the field variables $\phi$, their variations $\delta\phi$,
and the vector field $\xi^{\mu}$.

Now consider a black hole spacetime with a bifurcation (D-2) surface $\Sigma$ and a
Killing vector
\begin{equation}
\xi^{\mu}=t^{\mu}+\Omega^{(A)}_H\phi^{\mu}_{(A)}
\end{equation}
which vanishes on $\Sigma$. The vector $t^{\mu}$ is the stationary Killing vector with unit norm at infinity, $\psi_{(A)}^{\mu}$ are axial Killing vectors, and $\Omega_H^{(A)}$
is the angular velocity of the horizon. One should note that $\xi_{\mu}$ is defined with
respect to the connection, $\tilde{\nabla}_{\mu}$, that is compatible with $g_{\mu\nu}$
(i.e. $(L_{\xi}g)_{\mu\nu}=\tilde{\nabla}_{\mu}\xi_{\nu}+\tilde{\nabla}_{\nu}\xi_
{\mu}=0$).

The expression
\begin{equation}
\delta\int_{\Sigma}{\bf Q}=
\delta E-\Omega_{H}^{(A)}\delta J_{(A)}\; ,
\label{firstlaw}
\end{equation}
where
$E$ is the canonical energy and $J_{(A)}$ is the canonical angular momentum is also valid in the Palatini formalism.
To be able to interpret the left hand side of (\ref{firstlaw}) as being proportional to the entropy the dependence of ${\bf Q}$ on $\xi^{\mu}$ must be
eliminated. As in the metric formulation all second order and higher derivatives of $\xi^
{\mu}$ can be reduced to the first order derivatives $\nabla_{\mu}\xi_{\nu}$. If $\nabla_
{\mu}$ is compatible with some metric $h_{\mu\nu}$ then $\nabla_{\mu}\xi_
{\nu}=\kappa\epsilon_{\mu\nu}$ where $\kappa$ is the surface gravity and $\epsilon_
{\mu\nu}$ is the binormal to $\Sigma$ (defined relative to $h_{\mu\nu}$). Since
$\xi^{\mu}=0$ on $\Sigma$ the dependence of ${\bf Q}$ on $\xi^{\mu}$ can be
eliminated. Defining $\tilde{\xi}^{\mu}=\kappa\xi^{\mu}$ we have $\delta {\bf Q}
=\kappa\delta\tilde{{\bf Q}}$ and the entropy of a black hole is given by
\begin{equation}
S=2\pi\int_{\Sigma}\tilde{{\bf Q}}\; .
\end{equation}
\section{Entropy in f(R) Theories}
Consider Lagrangians of the form ${\bf L}=L{\bf \epsilon}$ in $D>2$ with
\begin{equation}
L=\frac{1}{16\pi G}f(R) +L_M\; ,
\end{equation}
where $R_{\mu\nu}=R^{\alpha}_{\;\;\mu\alpha\nu}$, $R=g^{\mu\nu}R_{\mu\nu}$,
and $L_M$ is the matter Lagrangian.
Varying the action with respect to $g^{\mu\nu}$ gives
\begin{equation}
f'(R)R_{(\mu\nu)}-\frac{1}{2}f(R)g_{\mu\nu}=8\pi GT_{\mu\nu}\; ,
\label{FE1}
\end{equation}
where $f'=df/dR$ and $R_{(\mu\nu)}$ is the symmetric part of the Ricci tensor.
Varying the action with respect to $\Gamma^{\alpha}_{\mu\nu}$ and simplifying gives
\begin{equation}
\nabla_{\alpha}\left[\sqrt{-g}f'g^{\mu\nu}\right]=0\; .
\label{eqn1}
\end{equation}
Defining $h_{\mu\nu}$ by $\sqrt{-h}h^{\mu\nu}=f'\sqrt{-g}g^{\mu\nu}$ gives \begin{equation}
\nabla_{\alpha}[\sqrt{-h}h^{\mu\nu}]=0\; ,
\end{equation}
so that the connection is compatible with $h_
{\mu\nu}$. Contracting (\ref{FE1}) over $\mu$ and $\nu$ gives
\begin{equation}
Rf'(R)-\frac{D}{2}f(R)=8\pi GT\; ,
\label{cont}
\end{equation}
which allows us to write $R=R(T)$.

First consider the entropy of vacuum spacetimes. In this case (\ref{cont}) implies that $R$ is a constant (unless $f\propto R^{D/2}$) and that $\nabla_{\mu}$ is also
compatible with $g_{\mu\nu}$.
 Using (\ref{current})
and (\ref{Gamma}) gives
\begin{equation}
J_{\mu_2...\mu_D}=\frac{1}{8\pi G}\left[\nabla_{\lambda}\left(f'\nabla^{[\lambda}\xi^
{\alpha]}\right)+\left(f'R^{\alpha}_{\;\;\;\lambda}
-\frac{1}{2}\delta^{\alpha}_{\;\;\;\lambda}f\right)\xi^{\lambda}\right]\epsilon_
{\alpha\mu_2...\mu_D}\; ,
\label{J1}
\end{equation}
where indices on $\nabla_{\mu}$ and $R_{\mu\nu}$ have been raised with $g^{\mu\nu}$.
It is also important to note that $\xi_{\mu}$ is raised by the metric $g^{\mu\nu}$, since the
Killing vector $\xi^{\mu}$ is used to determine the energy and momentum of the
spacetime with metric $g_{\mu\nu}$.  Thus, when the
field equations are satisfied
\begin{equation}
J_{\mu_2...\mu_D}=\frac{f'}{8\pi G}\nabla_{\lambda}\left(\nabla
^{[\lambda}\xi^{\alpha]}\right)\epsilon_
{\alpha\mu_2...\mu_D}\; .
\label{J}
\end{equation}
 The Noether potential ${\bf Q}$ can then be written as
\begin{equation}
Q_{\mu_3...\mu_D}=-\frac{f'}{16\pi G}\left(\nabla^{\alpha}\xi^{\beta}
\right)\epsilon_{\alpha\beta\mu_3...\mu_D}\; .
\label{Q}
\end{equation}
This is the same expression as one obtains in general relativity except for the constant factor $f'$. The entropy of a black hole is therefore given by
\begin{equation}
S_{BH}=f'\left(\frac{A}{4G}\right)\; .
\label{entropy}
\end{equation}

Now consider spacetimes with matter.
Assuming that $L_M$ is independent of the connection it is
easy to see that the matter Lagrangian will contribute the same terms to
(\ref{J1}) as it does in the metric approach. Forms of matter such as electromagnetic
fields and scalar fields only contribute terms that, with the curvature terms,
vanish when the field equations are satisfied. Thus, when the field equations are satisfied the Noether current is given by
\begin{equation}
J_{\mu_2...\mu_D}=\frac{1}{8\pi G}\nabla_{\lambda}\left[\nabla_{\sigma}
\left(h^{\sigma[\lambda}\xi^{\alpha]}\right)\right]
\epsilon_
{\alpha\mu_2...\mu_D}^{(h)}\; ,
\label{Jh}
\end{equation}
where ${\bf \epsilon}^{(h)}$ is the volume form associated with the metric
$h_{\mu\nu}$.
From
\begin{equation}
h_{\mu\nu}=(f')^pg_{\mu\nu}\; ,
\end{equation}
where $p=2/(D-2)$, the Noether charge can be written as
\begin{equation}
Q_{\mu_3...\mu_D}=-\frac{1}{16\pi G}f'g^{\alpha\sigma}\nabla_{\sigma}\left[
(f')^{p}h^{\beta\lambda}\xi_{\lambda}\right]\epsilon_{\alpha\beta\mu_3...\mu_D}\; .
\end{equation}
This can be expanded to obtain
\begin{equation}
Q_{\mu_3...\mu_D}=-\frac{1}{16\pi G}\left[f'g^{\alpha\sigma}\left(g^{\beta\lambda}
\nabla_{\sigma}\xi_{\lambda}+(f')^{-p}\nabla_{\alpha}[(f')^{p}]\xi^{\beta}\right)\right]
\epsilon_{\alpha\beta\mu_3...\mu_D}.
\label{Qf}
\end{equation}
Since $\xi^{\beta}=0$ on $\Sigma$ the Noether potential ${\bf Q}$ reduces to (\ref{Q}) on $\Sigma$, with $\nabla_{\sigma}$ compatible with $h_{\mu\nu}$
not with $g_{\mu\nu}$. However, ${\bf Q}$ is independent of
the connection, so that
the entropy is given by (\ref{entropy}) if $f'$ is constant on $\Sigma$.

Before concluding that the entropy of a black hole is $f'$ times one-quarter
of its area it is important to consider the coupling of matter to the gravitational
field. For simplicity consider conformally invariant matter with $T=0$. From
(\ref{cont}) we see that $R$ is a constant and the connection is compatible with $g_{\mu\nu}$.
The field equation (\ref{FE1}) can be written as
\begin{equation}
G_{\mu\nu}(g)=8\pi\left(\frac{G}{f'}\right)T_{\mu\nu}-
\left(\frac{D-2}{2D}\right) R g_{\mu\nu}\; ,
\label{Fe1a}
\end{equation}
where $G_{\mu\nu}(g)$ denotes the Einstein tensor constructed
from $g_{\mu\nu}$. The effective Newton's constant is therefore given by
\begin{equation}
G_N=\frac{G}{f'}
\label{G}
\end{equation}
and the entropy is given by
\begin{equation}
S_{BH}=\frac{A}{4G_N}\; .
\end{equation}
Thus, the entropy of a black hole in a spacetime with conformally invariant
matter is one-quarter of its area. This will also hold in vacuum spacetimes.

Another way to look at this is to consider the ``canonical" energy
\begin{equation}
E=\int_{\infty}\left({\bf Q}[t]-t\cdot {\bf B}\right)
\end{equation}
where
\begin{equation}
\delta\int_{\infty}\xi\cdot {\bf B}=\int_{\infty}\xi\cdot{\bf \Theta}\; .
\end{equation}
Iyer and Wald have shown \cite{Wa2} that $E$ is the ADM mass for general relativity.
If $G$ is taken to be Newton's constant then ${\bf Q}_f=f'{\bf Q}_{GR}$ and $E_f=f'E_{GR}$,
where ${\bf Q}_f$ refers to ${\bf Q}$ in the $f(R)$ theory and ${\bf Q}_{GR}$ refers to ${\bf Q}$
in general relativity. However, vacuum $f(R)$ theories are equivalent to Einstein's
theory (plus a possible cosmological constant)\cite{Fe1}.
Thus, it makes sense to
define Newton's constant by (\ref{G}), so that the energies coincide.

Now consider spacetimes containing matter that is not conformally invariant.
Equation (\ref{Fe1a}) holds, with additional terms on the right hand side
due to the fact that $\nabla_{\mu}$ is no longer compatible with $g_{\mu\nu}$.
In addition $f'$ is no longer constant.
If the energy-momentum tensor of the non-conformally invariant
matter vanishes at infinity $f'$ will approach a constant $f'_{\infty}$ at
large distances from the black hole. In this case Newton's constant will
be taken to be $G/f'_{\infty}$ and the entropy will be given by
\begin{equation}
S=\left(\frac{f'_{\Sigma}}{f'_{\infty}}\right)\frac{A}{4G_N}.
\end{equation}
if $f'$ is constant on $\Sigma$.
Thus, the entropy of a black hole is generally not equal to one-quarter
of its area (in Planck units) if non-conformally invariant matter is present.
\section{Additional Examples}
Consider Lagrangians of the form ${\bf L}=L{\bf
\epsilon}$ with
\begin{equation}
L=\frac{1}{16\pi G}\sqrt{-g}f(g^{\mu\nu},R_{(\mu\nu)})\; .
\end{equation}
Varying the action with respect to $g_{\mu\nu}$ gives
\begin{equation}
f_{\mu\nu}-\frac{1}{2}fg_{\mu\nu}=0\; ,
\label{32}
\end{equation}
where $f_{\mu\nu}=\partial f/\partial g^{\mu\nu}$. Varying the
action with respect to $\Gamma^{\alpha}_{\mu\nu}$ gives
\begin{equation}
\nabla\left[\sqrt{-g}\Im^{\mu\nu}\right]=0
\end{equation}
where $\Im^{\mu\nu}=\partial f/\partial R_{(\mu\nu)}$.
Now define
\begin{equation}
\sqrt{-h}h^{\mu\nu}=\sqrt{-g}\Im^{\mu\nu}
\label{h}
\end{equation}
and the connection is again
compatible with $h_{\mu\nu}$. It is easy to see that
\begin{equation}
g^{\mu\alpha}f_{\alpha\nu}=\Im^{\mu\alpha}R_{(\alpha\nu)}
\label{equiv}
\end{equation}
and that ${\bf J}$ is given by (\ref{Jh}) when the field equations
are satisfied. It has been shown \cite{Ta1,Al1,Bo1} that the field equations in this theory
are equivalent to Einstein's theory plus a possible cosmological constant
(except on a set of measure zero). Since $\Im^{\mu\nu}$ is constructed from
$g^{\mu\nu}$ and $R_{(\mu\nu)}$ it must be of the form
\begin{equation}
\Im^{\mu\nu}=\lambda g^{\mu\nu}
\label{conv}
\end{equation}
when the field equations are satisfied, where $\lambda$ is a constant which I will assume is nonzero. From (\ref{h}) and (\ref{conv})
we have
\begin{equation}
\sqrt{-h}h^{\mu\nu}=\lambda\sqrt{-g}g^{\mu\nu}\; ,
\end{equation}
which implies that $S_{BH}=\lambda(A/4G)$. Including matter in (\ref{32}) and using
(\ref{equiv}) and (\ref{conv})
shows that the effective gravitational coupling is $G_N=G/\lambda$. This
can also be seen by equating the energy in these theories with the energy
in general relativity. Thus, $S_{BH}=A/4G_N$.

As a final example consider the Lagrangian ${\bf L}=
L{\bf \epsilon}$ where
\begin{equation}
L=e^{-2\phi}\left[\frac{1}{16\pi G}f(R)-\frac{1}{2}\alpha\nabla_{\mu}\phi
\nabla^{\mu}\phi\right]\; ,
\label{Lag}
\end{equation}
and $\alpha$ is a constant.
The field equations that follow from varying the metric and connection are
\begin{equation}
f'R_{(\mu\nu)}-\frac{1}{2}fg_{\mu\nu}=8\pi G\alpha \left[\nabla_{\mu}\phi
\nabla_{\nu}\phi-\frac{1}{2}g_{\mu\nu}(\nabla\phi)^2\right]
\label{dil1}
\end{equation}
and
\begin{equation}
\nabla_{\alpha}\left[\sqrt{-g}e^{-2\phi}f'g^{\mu\nu}\right]=0\; .
\label{dil2}
\end{equation}
Defining
$\sqrt{-h}h^{\mu\nu}=\sqrt{-g}e^{-2\phi}f'g^{\mu\nu}$ gives $\nabla_{\alpha}(\sqrt{-h}
h^{\mu\nu})=0$.
The Noether charge is given by
\begin{equation}
Q_{\mu_3...\mu_D}=-\frac{1}{16\pi G}e^{-2\phi}f'
g^{\alpha\sigma}
\left(\nabla_{\sigma}\xi^{\beta}\right)
\epsilon_{\alpha\beta\mu_3...\mu_D}\; ,
\end{equation}
and the entropy is
\begin{equation}
S_{BH}=e^{-2\phi^{(\Sigma)}}f'_{\Sigma}\left(\frac{A}{4G}\right)\; ,
\end{equation}
where $\phi^{(\Sigma)}$ and $f'_{\Sigma}$ are evaluated on $\Sigma$ (assuming that they are constant on $\Sigma$).
However, from (\ref{dil1}) one can define
$G_N=G/f'_{\infty}$ and this gives
\begin{equation}
S_{BH}=e^{-2\phi^{(\Sigma)}}\frac{f'_{\Sigma}}{f'_{\infty}}
\left(\frac{A}{4G_N}\right)
\end{equation}
for the entropy of a black hole. In string theory $\alpha\propto 1/G$, so that equation
(\ref{dil1}) is actually independent of $G$. One can still define $G_N=G/f'_{\infty}$,
since the coefficient of $R_{(\mu\nu)}$ is $f'/G$.

\section{Conclusion}
In this paper I examined black hole entropy in generalized theories of gravity
in the Palatini formalism using the Noether charge approach. It was shown
that the entropy of a black hole in $f(R)$ theories with conformally invariant matter is given by
\begin{equation}
S_{BH}=\frac{A}{4G_N}\; ,
\end{equation}
where $G_N=G/f'$ is the effective Newtonian constant for the theory and $f'$
is a constant. For theories with matter that is not conformally invariant
$f'$ is not a constant, $G_N=G/f'_{\infty}$ and
\begin{equation}
S=\left(\frac{f'_{\Sigma}}{f'_{\infty}}\right)\frac{A}{4G_N}\; ,
\end{equation}
 where $f'_{(\Sigma)}$ is
the value of $f'$ evaluated on the bifurcation surface and $f'_{\infty}$
is its value at spatial infinity.
Vacuum theories with Lagrangians depending on arbitrary functions of the
symmetric part of the Ricci tensor were examined. The field equations
of these theories are equivalent to Einstein's theory plus a possible
cosmological constant and the entropy was shown to be one-quarter of the
surface area. Finally generalized dilaton theories of the
form
\begin{equation}
L=e^{-2\phi}\left[\frac{1}{16\pi G}f(R)-\frac{1}{2}\alpha\nabla_{\mu}\phi
\nabla^{\mu}\phi\right]
\end{equation}
were considered and the entropy of a black hole was shown to be
\begin{equation}
S_{BH}=e^{-2\phi^{(\Sigma)}}\frac{f'_{\Sigma}}{f'_{\infty}}
\left(\frac{A}{4G_N}\right)\; ,
\end{equation}
where $\phi^{(\Sigma)}$ and $f'_{\Sigma}$ are evaluated on $\Sigma$ (assuming that they are constant on $\Sigma$)
and $f'_{\infty}$ is the value of $f'$ at infinity.

\section*{Acknowledgements}
I would like to thank Bill Unruh for helpful comments. This research was
supported by the  Natural Sciences and Engineering Research
Council of Canada.


\begin{thebibliography}{99}
\bibitem{1}
S.Capozziello, V.F. Cardone, S. Carloni, A. Troisi, Int. J. Mod. Phys. D\textbf{12},
1969 (2003); S.M. Carroll, V. Duvvuri, M. Trodden, and M.S. Turner, Phys. Rev. D\textbf{70},
043528 (2004); D.N. Vollick, Phys. Rev. D\textbf{68}, 063510 (2003)
\bibitem{others 1}
S. Nojiri and S. D. Odintsov, Phys. Lett. B{\bf 576}, 5 (2003);
A.D. Dolgov and M. Kawasaki, Phys. Lett.
B{\bf 573}, 1 (2003);
X.H. Meng and P. Wang, Class. Quant. Grav. {\bf 20}, 4949 (2003);
S. Nojiri and S.D. Odintsov, Phys. Rev. D{\bf 68}, 123512 (2003);
X.H. Meng and P. Wang, Class. Quant. Grav. {\bf 21}, 951 (2004);
S. Nojiri and S.D. Odintsov, Gen. Rel. Grav. {\bf 36} 1765 (2004);
X.H. Meng and  P. Wang, Phys. Lett. B{\bf 584}, 1 (2004);
S. Nojiri and S.D. Odintsov, Mod. Phys. Lett. A{\bf 19}, 627 (2004);
X.H. Meng, P. Wang, Class. Quant. Grav. {\bf 21}, 2029 (2004);
G. Allemandi, A. Borowiec, M. Francaviglia, Phys. Rev. D{\bf 70}, 043524 (2004);
G.M. Kremer and D.S.M. Alves, Phys. Rev. D{\bf 70}, 023503 (2004);
G. Allemandi, A. Borowiec and M. Francaviglia, Phys. Rev. D{\bf 70}, 103503
(2004);
M.C.B. Abdalla, S. Nojiri and S.D. Odintsov, Class. Quant. Grav. {\bf 22},
L35 (2005);
S.M. Carroll, A. De Felice, V. Duvvuri, D.A. Easson, M. Trodden and M.S. Turner Phys. Rev. D{\bf 71}, 063513 (2005);
S. Capozziello, V.F. Cardone and M. Francaviglia, Gen. Rel. Grav. {\bf 38},
711 (2006);
S. Capozziello, V.F. Cardone, M. Funaro and S. Andreon, Phys. Rev. D{\bf
70}, 123501 (2004);
S. Carloni, P.K.S. Dunsby and S. Capozziello and A. Troisi, Class. Quant. Grav.
{\bf 22}, 4839 (2005);
D.A. Easson, Int. J. Mod. Phys. A{\bf 19} 5343 (2004);
S. Capozziello, V.F. Cardone and A. Troisi, Phys. Rev. D{\bf 71}, 043503
(2005);
S. Das, N. Banerjee and N. Dadhich, Class. Quant. Grav. {\bf 23},
4159 (2006);
I. Brevik, O. Gorbunova and Y.A. Shaido, Int. J. Mod. Phys. D{\bf 14}
 1899 (2005);
S. Capozziello, V.F. Cardone, E. Elizalde, S. Nojiri and S.D. Odintsov,
Phys. Rev. D{\bf 73} 043512 (2006);
T. Koivisto and H. Kurki-Suonio, Class. Quant. Grav. {\bf 23}, 2355 (2006);
N.J. Poplawski, Class. Quant. Grav. {\bf 23}, 2011 (2006);
O. Mena, J. Santiago and J. Weller, Phys. Rev. Lett. {\bf 96}, 041103 (2006);
M. Amarzguioui, O. Elgaroy, D.F. Mota and T. Multamaki, Astron. Astrophys.
{\bf 454} 707 (2006);
T. Koivisto, Phys. Rev. D{\bf 73}, 083517 (2006);
L. Amendola, D. Polarski and S. Tsujikawa, Phys. Rev. Lett. {\bf 98}, 131302
(2007);
S. Capozziello, S. Nojiri, S.D. Odintsov and A. Troisi,
Phys. Lett. B{\bf 639}, 135 (2006);
S.M. Carroll, I. Sawicki, A. Silvestri and Mark Trodden ,
New J. Phys. {\bf 8}, 323 (2006);
J. A. R. Cembranos, Phys.Rev. D{\bf 73}, 064029 (2006);
S. Fay, R. Tavakol, S. Tsujikawa, Phys. Rev. D{\bf 75}, 063509 (2007);
I. Sawicki and W. Hu, Phys. Rev. D{\bf 75}, 127502 (2007);
S. Fay, S. Nesseris and L. Perivolaropoulos, Phys. Rev. D{\bf76}, 063504
(2007);
K. Uddin, J.E. Lidsey, R. Tavakol, Class. Quant. Grav. {\bf 24}, 3951 (2007);
M.S. Movahed, S. Baghram and S. Rahvar, Phys. Rev. D{\bf 76}, 044008 (2007);
T. Koivisto, Phys. Rev. D{\bf 76}, 043527 (2007);
Yong-Seon Song, H. Peiris and W. Hu, Phys. Rev. D{\bf 76}, 063517 (2007);
\bibitem{Wa1}
R. M. Wald, Phys. Rev. D\textbf{48}, R3427 (1993)
\bibitem{Wa2}
V. Iyer and R.M. Wald, Phys. Rev. D\textbf{50}, 846 (1994)
\bibitem{Vi1}
M. Visser, Phys. Rev. D\textbf{48}, 5697 (1993);
\bibitem{Ja1}
T. Jacobson, G. Kang and R.C. Myers, Phys. Rev. D\textbf{49}, 6587 (1994);
\bibitem{Wa3}
V. Iyer and R.M. Wald, Phys. Rev D\textbf{52}, 4430 (1995)
\bibitem{Br1}
F. Briscese and E. Elizade, arXiv:hep-th/0708.0432;
G. Cognola, E. Elizalde,  S. Nojiri, S.D. Odintsov and
S. Zerbini, J. Cosmol. Astropart. Phys. {\bf 0502}, 010 (2005);
G. Cognola, E. Elizalde,  S. Nojiri, S.D. Odintsov and
S. Zerbini, Phys. Rev. D{\bf 73}, 084007 (2006);
M. Akbar and Rong-Gen Cai, Phys. Lett. B\textbf{648}, 243 (2007);
H. Mohseni Sadjadi, arXiv:gr-qc/0709.2435;
A.Mukhopadhyay and T. Padmanabhan, Phys.Rev. D {\bf 74}, 124023 (2006);
A. Paranjape, S. Sarkar and T. Padmanabhan, Phys.Rev. D {\bf 74}, 104015 (2006);
P. Wang, Phys. Rev. D{\bf 72}, 024030 (2005);
M. Cvitan and S. Pallua, Phys. Rev. D{\bf 71}, 104032 (2005);
I. Brevik, S. Nojiri, S.D. Odintsov and L. Vanzo, Phys. Rev. D {\bf 70}, 043520 (2004);
J.I. Koga and K.I. Maeda, Phys. Rev. D{\bf 58}, 064020 (1998);
T. Jacobson, G. Kang and R.C. Myers, Phys. Rev. D{\bf 52}, 3518 (1995)
\bibitem{Wa4}
R.M. Wald, J. Math. Phys. {\bf 31}, 2378 (1990)
\bibitem{Fe1}
M. Ferraris, M. Francaviglia and I. Volovich, Class. Quant. Grav. \textbf{11}
, 1505 (1994)
\bibitem{Ta1}
V. Tapia and M. Ujevic, Class. Quant. Grav. \textbf{15}, 3719 (1998)
\bibitem{Al1}
G. Allemandi, A. Borowiec and M. Francaviglia, Phys. Rev. D\textbf{70},
103503 (2004)
\bibitem{Bo1}
A. Borowiec, M. Ferraris, M. Francaviglia and I. Volovich, Class. Quant.
Grav. \textbf{15}, 43 (1998)
\end{thebibliography}
\end{document}